\documentstyle[preprint,prb,aps]{revtex}
\begin{document}
\draft
\title
{Chiral Exponents of the Square Lattice Frustrated XY Model:\\ a Monte
Carlo Transfer Matrix calculation}

\author {Enzo Granato \\
Laborat\'orio Associado de Sensores e Materiais,\\
Instituto Nacional de Pesquisas Espaciais,\\
12.225 S\~ao Jos\'e dos Campos, S.P., Brazil.}
\author { M.P. Nightingale \\
Department of Physics,\\
University of Rhode Island,\\
Kingston, Rhode Island 02881.}

\maketitle \begin{abstract} Thermal and chiral critical exponents of
the fully frustrated XY model on a square-lattice are obtained from a
finite-size scaling analysis of the free energy of chiral domain
walls.  Data were obtained by extensive Monte Carlo transfer matrix
computations for infinite strips of widths up to $14$ lattice
spacings.  Two transfer matrices were implemented, one for each of the
two principal lattice directions.  The results of both are consistent,
but the critical exponents differ significantly from the pure Ising
values. This is in agreement with other recent Monte Carlo simulations.
Our results also support the identification of the critical behavior of
this model with that along the line of transitions of simultaneous
ordering or becoming critical of Ising and planar rotor degrees of
freedom in the XY-Ising model studied recently.  \end{abstract}


\section{Introduction}

There has been considerable interest, both experimentally and
theoretically, in phase transitions of two-dimensional, fully
frustrated XY models.  Most studies have been motivated by their
relevance to Josephson junction arrays in a magnetic field, where these
models describe the superconducting-to-normal transition at half a flux
quantum per plaquette \cite{Zant}, but there are also interesting
theoretical questions regarding the identification of the universality
class \cite{Villain,Teitel,Miyashita,Dlee,Himbergen,Berge,TK88,Grest,%
TK90,Granato91,LeeK,Jose,LeeG}.

In the standard XY model without frustration the relevant symmetry is
the continuous $U(1)$ symmetry which, as is well-known, is responsible
for a Kosterlitz-Thouless phase transition. The fully frustrated XY
model has a richer behavior with a low-temperature phase with critical
fluctuations associated with the $U(1)$ symmetry, accompanied by a
broken, discrete $Z_2$ symmetry. In principle, there are two different
ordering scenarios: ordering can take place in two stages via separate
XY and Ising transitions, or both symmetries can be broken or become
critical simultaneously, which yields a single transition, presumably
in a new universality class.

On a square lattice the  model can be defined by the Hamiltonian
\begin{equation}
 H  = - \sum_{<ij>}  J_{ij} \cos (\theta_i - \theta_j),
\label{eq.hamiltonian}
\end{equation}
where $J_{ij} =  J $ $(J>0)$  for
horizontal rows and $J_{ij} = \pm J$ for alternating vertical columns.
Owing to the presence of an odd number of antiferromagnetic bonds in
each plaquette, the model is fully frustrated.  This leads to a double
degeneracy of the ground state, which is of course accompanied by an
additional continuous degeneracy, a manifestation of the $U(1)$
symmetry.  One can introduce an Ising-like order parameter $\chi_p$,
the local chirality, which measures the sense of rotation of a spin of
unit length $\vec s = (\cos \theta , \sin \theta)$ along the sides of a
given plaquette $p$. In the ground state there is antiferromagnetic
order of the local chiralities ($\chi_p = \pm 1$).

Early Monte Carlo simulations led to the conclusion that the chiral
order parameter of the fully frustrated XY model has pure Ising model
critical exponents
\cite{Teitel,Miyashita,Dlee,Himbergen,Berge,TK88,Grest}, but more
recent \cite{TK90,Granato91,LeeK,Jose} estimates of the exponents have
differed significantly from the Ising values.  In particular,
predictions were made for the values of chiral critical exponents of
fully frustrated XY model.  These predictions are based on results for
the XY-Ising model \cite{Granato91,LeeG}, which is expected to describe
the critical behavior in these systems.  These ideas are supported by
recent Monte Carlo simulations \cite{LeeK,Jose}.  However, in view of
results that suggest pure Ising critical exponents, and in the absence
of precise agreement among the more recent estimates, the current state
of affairs is unsatisfactory from a computational point of view. The
additional numerical results presented in this paper may serve to help
settle the issue.

We report results of extensive numerical calculations of the chiral
domain wall free energy of the fully frustrated XY model in an infinite
strip geometry.  Because of the continuous nature of the phase
variables of this model, the transfer matrix is intractable for
numerically exact computation of its eigenvalues.  We therefore use the
Monte Carlo transfer matrix method \cite{NightingaleBloetePRL88} to
obtain the free energy from the largest eigenvalue of the transfer
matrix.  Exploiting the anisotropy of the hamiltonian
(\ref{eq.hamiltonian}) we use two different implementations, a
``horizontal" ({\it i.e.} column-to-column) and ``vertical" ({\it i.e.}
row-to-row) transfer matrix. These approaches yield results in
reasonable mutual agreement.  From a finite-size scaling analysis of
data for strips of widths up to $14$ lattice spacings, we have
estimated  the thermal  exponent $\nu$, the exponent $\eta$ associated
with the Ising-like order-order correlation function and the critical
temperature $T_{\rm c}$.  In particular, the result obtained for $\nu$
is quite insensitive of the estimate of $T_{\rm c}$, as will be
discussed in more detail below.  We summarize our estimates (with error
estimates in parentheses):  $\nu = 0.80(4)$, $\eta = 0.38(3)$ and
$T_{\rm c} = 0.454(4)$.  Within the errors, these numbers are in
agreement, with a Monte Carlo simulation using similar finite-size
scaling analysis \cite{LeeK} and they are also consistent with other
Monte Carlo simulations \cite{Jose}. However, our value of $\nu$
disagrees with the estimate of $\nu = 1$ by Thijssen and Knops
\cite{TK90}, who also used the Monte Carlo transfer matrix method, but
a different method to estimate $\nu$; their deviating estimate is
likely to be an artifact of their fitting procedure \cite{TKfootnote}.
On the other hand, our value of $\eta$ is in very good agreement with
their estimate.  Moreover, our results appear to support the
identification of the critical behavior of the fully frustrated XY
model with the line of single transitions in XY-Ising model studied
recently \cite{Granato91,LeeG}.

\section{Monte Carlo Transfer Matrix}

The Monte Carlo transfer matrix method introduced by Nightingale and
Bl\"ote \cite{NightingaleBloetePRL88}, is particularly useful when, as
is the case in the model under consideration, the continuous nature of
the spin variables does not allow a numerically exact diagonalization
of the transfer matrix. The method is a stochastic version of the
well-known power method of calculating the dominant eigenvalue of a
matrix or integral kernel.  More specifically, the dominant eigenvector
$\psi^{0}$ is approximated by a Monte Carlo time average over weighted
walkers representing row (or column) configurations.  The basic idea is
that the (weighted) frequency of occurrence of a particular row (or
column) configuration, say $S=(\vec s_1, \vec s_2,\cdots, \vec s_L)$,
is proportional to the magnitude of the corresponding component
$\psi^{0}_S$ of the dominant eigenvector.

The key elements of the algorithm are the following. Walkers are
generated in subsequent generations labeled by an index $t$.  The
generation at time $t$ consists of a sequence of a number of $r_t$
walkers
$[(S_{1,t},w_{1,t}),(S_{2,t},w_{2,t}),\cdots,(S_{r_t,t},w_{r_t,t})]$,
where the $S_{i,t}$ are row (or column) configurations of the form $S$
introduced above, and the $w_{i,t}>0$ are statistical weights.  This
sequence represents a (generally extremely) sparse vector $\psi$ with
components
\begin{equation}
 \psi_S=\sum_{i=1}^{r_t} \delta_{S,S_i} w_{i,t},
\end{equation}
where $\delta$ is the Kronecker $\delta$-function.  One can write
\begin{equation}
 \psi'_{S'}=\sum_S T(S'|S) \psi_{S}=\sum_S P(S'|S) D_S
\psi_{S},
\label{eq.psi'}
\end{equation}
where $D_S=\sum_{S'} T(S'|S)$
and $P(S'|S)=T(S'|S)/D_S$.  Since by construction $P$ is a stochastic
matrix, multiplication by the transfer matrix of a vector of the form
$\psi$ can be implemented as a stochastic process with transitions from
$S$ to $S^\prime$ with probability $P(S'|S)$.  That is, to update
generation $t$ to $t+1$, new walker states are sampled with probability
$P(S_{t+1}|S_t)$, while in each transition the weight of walker $i$ is
updated to $w_{i,t+1} = D_S w_{i,t}/c_{t+1}$.  For reasons of
efficiency the weights are kept close to unity by duplicating walkers
with great statistical weights and by eliminating walkers with low
weights. To ensure that $r_{t+1}$ remains close to its initial value
$r_0$, one can choose $c_{t+1}=\lambda_t r_t/r_0$, where $\lambda_t$ is
a moving or cumulative average estimate of the dominant transfer matrix
eigenvalue.  Assuming that the generation counter $t$ is reset to unity
upon equilibration, the largest eigenvalue can be estimated from a
sequence of $T$ generations as
\begin{equation}
 \lambda = \sum_{t=1}^T c_{t+1} W_{t+1} / \sum_{t=1}^T W_t,
\label{eq.lambda}
\end{equation}
where $W_{t'}=\sum_{i} w_{i,t'}$ denotes the total statistical weight
of all walkers of generation $t'$.  It should be noted that this
estimator of the dominant eigenvalue suffers from a bias due to the
correlations of the population control constants $c_t$ with the weights
$W_t$.  These correlations have the effect of suppressing large
contributions to the estimator (\ref{eq.lambda}) and enhancing small
ones \cite{UmrigarNR}.  As a result, the estimator tends to
underestimate the exact transfer matrix eigenvalue, but this effect can
be reduced by choosing a large target number of walkers, $r_0$.
Further algorithmic details, such as how to correct for this bias
without increasing the average population size and an alternative,
sometimes better choice of the population control constant $c_t$ can be
found in
Refs.~\onlinecite{Hetherington84,UmrigarNR,NightingaleBloteSpin1,%
NightingaleRev90}.  In applying this method to the fully frustrated XY
model, we have performed extensive calculations using, typically, $r_0
= 30,000$ walkers and $150,000$ Monte Carlo steps which corresponds to
$4.5 \times 10^9$ attempts per site.

To apply this Monte Carlo method it is necessary to sample
configurations $S'$ from the distribution $P(S'|S)$ for arbitrary $S$
and to evaluate the quantity $D_S$ at each elementary Monte Carlo
step.  One way to do this is to factor the transfer matrix in a way
that amounts to building up the square lattice by, {\it e.g.}, first
adding bonds perpendicular to the transfer direction and then adding
new sites and bonds along the transfer direction.  The algorithm given
above is then applied to both factors in succession.  The first of
these operations corresponds to multiplication by a diagonal matrix.
This leaves the row (or column) configurations unchanged while the
weight factors are calculated trivially:  they consist of one term
each.  The second matrix factor, which adds $L$ new sites, has a direct
product form. Sampling and reweighting problems in this case reduce to
simple one-site problems.  In principle, this method works, but it
yields an inefficient Monte Carlo process: the simple factorization
generates row (or column) configurations with variables that are too
weakly correlated in the sense that, as it is added to the lattice,
each variable is sampled independently from a distribution with direct
correlations only to one other nearest-neighbor variable.  True
many-site correlations are subsequently recovered via multiplication by
strongly fluctuating weight factors. This, however, causes strong
fluctuations in the number of walkers, which results in frequent
elimination and duplication of walkers and enhances correlations
between walkers within each generation, which reduces the overall Monte
Carlo sampling efficiency.

The alternative we have used in this paper requires helical boundary
conditions \cite{NightingaleBloetePRL88}.  In that case the lattice can
be constructed by repetition of identical elementary steps, each of
which adds one site at a time.  This implies that the states $S'$ and
$S$ in the transition matrix $P(S'|S)$ consist of lattice sites all of
which coincide with the exception of one site.  Consequently only a
small number of states $S'$ can be reached from any given state $S$,
{\it i.e.}, $P$ is sparse in this case.  The degrees of freedom that
are added will have direct correlations with those at {\it two}
neighboring lattice sites, one in the horizontal and one in the
vertical direction for the square lattice.  In principle, given an
approximate dominant eigenvector, one can transform the transfer matrix
and incorporate even more correlations from the onset, and thus reduce
the fluctuations of the weights and eigenvalue estimates
\cite{NightingaleRev90,NightingaleCaflisch88,NightingaleSuper88}.  We
will return to this later on.

Use of helical boundary yields a more efficient Monte Carlo process,
but the method has the disadvantage of producing a lattice with a
surface that has a step-defect, as is unavoidable when one cuts across
a screw.  One would expect the presence of this defect to lead to
unnecessary corrections-to-scaling, which may adversely affect
finite-size convergence, but in practice this appears not to be a
serious problem.

In fact, the method used in this paper requires use of a transfer
matrix of somewhat more complicated form than the one of Eq.
\ref{eq.psi'}, in that this equation has to be replaced by
\begin{equation}
 \psi''_{S''}=\sum_{S',S} T^{(1)}(S''|S')T^{(2)}(S'|S)
\psi_{S}.  \label{eq.psi''}
\end{equation}
Although this is no real
complication --one can simply devise a stochastic process with steps
alternating in correspondence to the two matrices $T^{(1)}$ and
$T^{(2)}$-- it has prevented our use of approximate trial eigenvectors
to reduce the noise of the stochastic process. In principle, this
variance reduction scheme works as follows.  Eq. \ref{eq.psi'} can be
replaced by the equivalent equation
\begin{equation}
\hat{\psi}'_{S'}=\sum_S \hat{T}(S'|S) \hat{\psi}_{S} \label{eq.psihat'}
\end{equation}
where $\hat{T}(S'|S)=\gamma_{S'} T(S'|S)/ \gamma_{S}$
and $\psi$ and $\psi'$ are similarly similarity transformed.  As long
as the components of $\gamma$ are all of one sign and do not vanish,
the same Monte Carlo method can be applied to $\hat{T}$ instead of
$T$.  This approach satisfies the following zero-variance principle: in
the ideal limit where $\gamma$ is the dominant left eigenvector of $T$,
the largest eigenvalue of $T$ can be estimated with vanishing
statistical error.  More realistically, as this limit is approached the
variance of the Monte Carlo process decreases.  In practical
applications, one chooses a trial vector $\gamma(p_1,p_2,\cdots)$ which
depends on variational parameters $p_i$. These are optimized by
minimization of the variance over $S$ of $\sum_{S'} \gamma_{S'}
T(S'|S)/\gamma_S$ with states selected with probability proportional to
$\gamma_S^2$.  This minimization is accomplished approximately by
minimizing the variance over a relatively small set of states $S$
generated by Monte Carlo \cite{UmrigarWilsonWilkins}.  For a transfer
matrix of the form of Eq. \ref{eq.psi''} the same process is still
possible in principle, but it is more complicated.  In fact there are
two alternatives.  One can define a Kronecker-product-like, two-site
transfer matrix $T^{(12)}(S'',S'|S',S)=T^{(1)}(S''|S')T^{(2)}(S'|S)$
and base the Monte Carlo process on the derived stochastic matrix
$P(S'',S'|S',S) \propto \gamma^{(12)}_{S'',S'} T^{(12)}(S'',S'|S',S)/
\gamma^{(12)}_{S',S}$, where ideally $\gamma^{(12)}$ is the dominant
left eigenvector of $T^{(12)}$.  We note that the pairs $(S'',S')$ and
$(S',S)$ differ at two lattice sites, and therefore this approach
requires simultaneous sampling of two site variables, which renders the
algorithm unnecessarily slow.  The alternative is to employ --as we
have done in this paper-- a matrix product of two single-site transfer
matrices rather than a single two-site transfer matrix, but this
approach requires two trial vectors.  In terms of these, one defines
$\hat{T}^{(1)}(S'|S)=\gamma^{(1)}_{S'} T^{(1)}(S'|S) / \gamma^{(2)}_S$,
and $\hat{T}^{(2)}_{S',S}=\gamma^{(2)}_{S'} T^{(2)}(S'|S)/
\gamma^{(1)}_S$.  It is straightforward to construct a transformed
process which again has a zero variance principle.  This time it
requires that $\gamma^{(1)}$ be the dominant left eigenvector of the
product matrix $T^{(1)}T^{(2)}$ and that $\gamma^{(2)}$ be the dominant
left eigenvector of $T^{(2)}T^{(1)}$.  Again adjustable parameters of
$\gamma^{(1)}$ and $\gamma^{(2)}$ can be chosen by minimization of the
(appropriately weighted) sum of the variances of $\sum_{S'} \gamma_{S'}
T(S'|S)/\gamma_S$ and $\sum_{S'} \gamma_{S'} T(S'|S)/\gamma_S$.  In
contrast with the alternative of a two-site transfer matrix, the
presence of two matrices slows down the algorithm only in the initial
stage of parameter optimization with this approach.

\section{Chiral Domain Wall Free Energy}

For an infinite strip of width $L$, the reduced free energy $f$ per
lattice site can be obtained from the largest transfer matrix
eigenvalue, $\lambda(L,K)$, via the relation $f=-\log \lambda$, where
$K=J/k_{\rm B} T$ is the reduced coupling constant.  For any given $L$
this quantity, $f$, depends on the choice of boundary conditions.  By
suitable choice of the latter, as specified in detail below, the chiral
domain wall free energy can be obtained from the free energy
difference, denoted by $\Delta f$.  For finite-size analysis, a
convenient quantity is the domain wall energy per $L$ lattice units of
length:
\begin{equation}
 \Delta F(K,L) = L^2 \Delta f(K,L).
\end{equation}

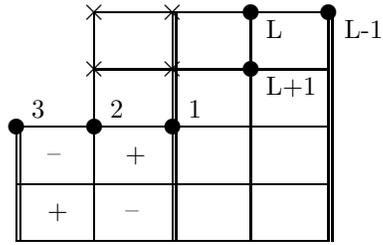
\begin{figure} \caption{Schematic of the process of building up the
infinite strip using a transfer matrix. This can be implemented using a
Monte Carlo transfer matrix along the horizontal direction with double
helical boundary conditions (a) or along the vertical direction with
single (b) or double (c) helical boundary conditions. At each step a
new configuration is obtained from the previous one, indicated by
$\bullet$, by adding a new spin $\times$ through a Monte Carlo process
and relabeling the sites. The $\pm $ indicates the antiferromagnetic
pattern of the chirality in the ground state.} \label{fig.transfer}
\end{figure}

Since the fully frustrated XY model is  spatially anisotropic, one can
devise two different types of boundary conditions to compute the domain
wall free energy. These two types are associated with two different
transfer matrices obtained by choosing the transfer direction to be
either horizontal or vertical, as shown in Fig.~\ref{fig.transfer}.  If
the transfer direction is horizontal (Fig.~\ref{fig.transfer}.a), one
is forced to use helical boundary conditions with a pitch of two
measured in lattice units, so as to match up the vertical
antiferromagnetic bonds, which have a periodicity of two in the
transfer direction. This is the construction used in
Ref.~\onlinecite{TK90}.  As indicated in Fig.~\ref{fig.transfer}.a,
only strips with $L$ even will match the antiferromagnetic pattern of
the local chiralities $\chi_p$ in the ground state; for strips with odd
$L$ the boundary conditions will introduce a chiral domain wall along
the infinite horizontal direction.  Calculation of the domain wall
energy requires that the boundary conditions be varied at constant
$L$.  With the horizontal transfer matrix this can be done
approximately only, and we chose to use the difference of the free
energy computed directly for $L$, and the free energy obtained by linear
interpolation between sizes $L-1$ and $L+1$.  This still leaves two
possibilities, depending on whether $L$ is odd or even.  As an
alternative we also performed calculations using a vertical transfer
matrix illustrated in Fig.~\ref{fig.transfer}.  In this case one has a
choice between helical boundary conditions with a pitch of one or two
lattice units. For each value of $L$ precisely one of these two
boundary conditions will force the presence of a domain wall and this
offers a convenient way to determine the chiral domain free energy
without interpolation.

Calculations of the free energy using the horizontal and vertical
transfer matrices were performed as a function of strip width in a very
narrow range of couplings $\Delta K = 0.04$ around the estimate of the
critical coupling obtained by Monte Carlo simulations of Lee {\it et
al.}\cite{LeeK}.  Note that the range we use is about ten times smaller
than the range used in Ref.~\onlinecite {TK90}.  The data for the
chiral domain wall free energy using both implementations of the
transfer matrix are shown in Figs.~\ref{fig.DeltaFvsLh} and
\ref{fig.DeltaFvsLv}.

\section{Finite-size Scaling Analysis}

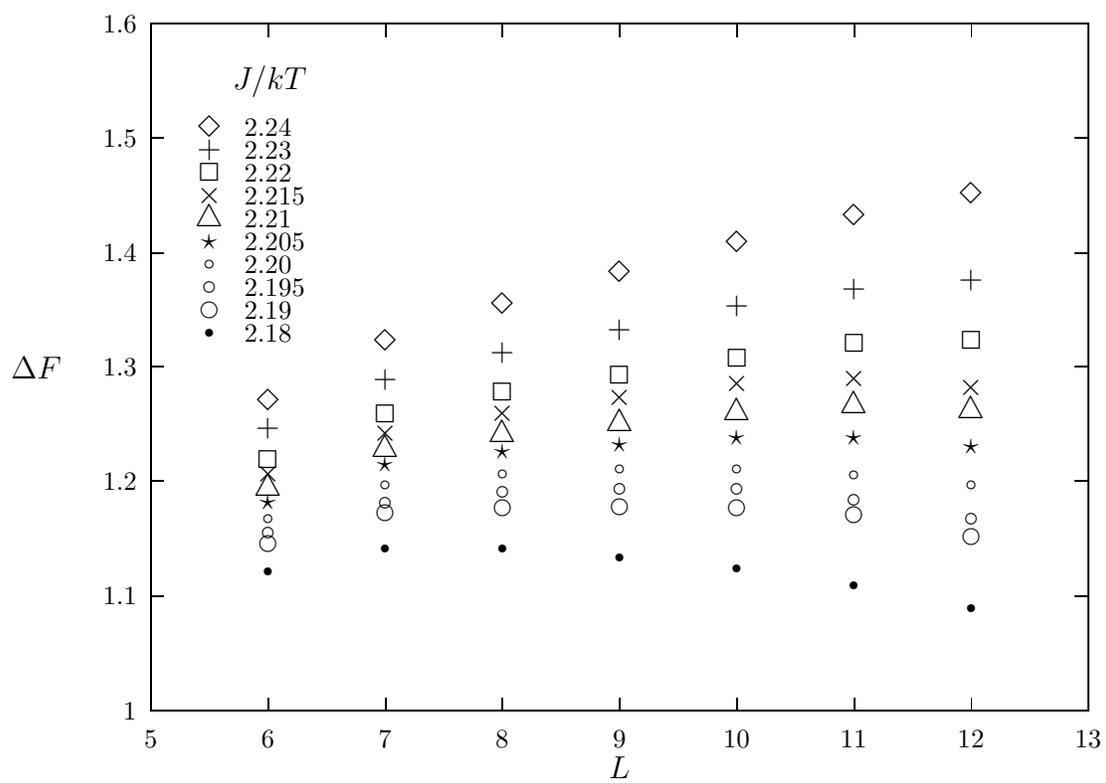
\begin{figure} \caption{Data for the chiral domain wall free energy
$\Delta F(K,L)$, obtained from horizontal Monte Carlo transfer matrix
calculations, in a small range around $T_{\rm c}$. }
\label{fig.DeltaFvsLh} \end{figure}

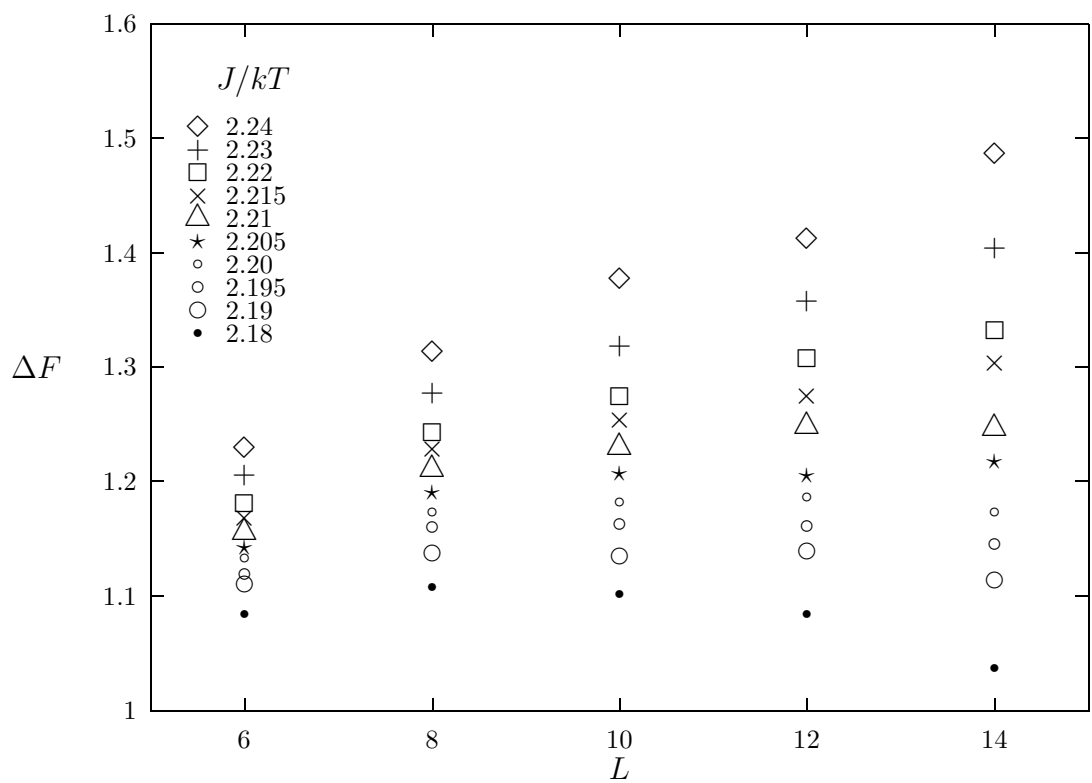
\begin{figure} \caption{Chiral domain wall free energy $\Delta F(K,L)$,
obtained from  vertical Monte Carlo transfer matrix calculations, for
the same couplings $K$ of Fig.~\protect\ref{fig.DeltaFvsLh}.}
\label{fig.DeltaFvsLv} \end{figure}

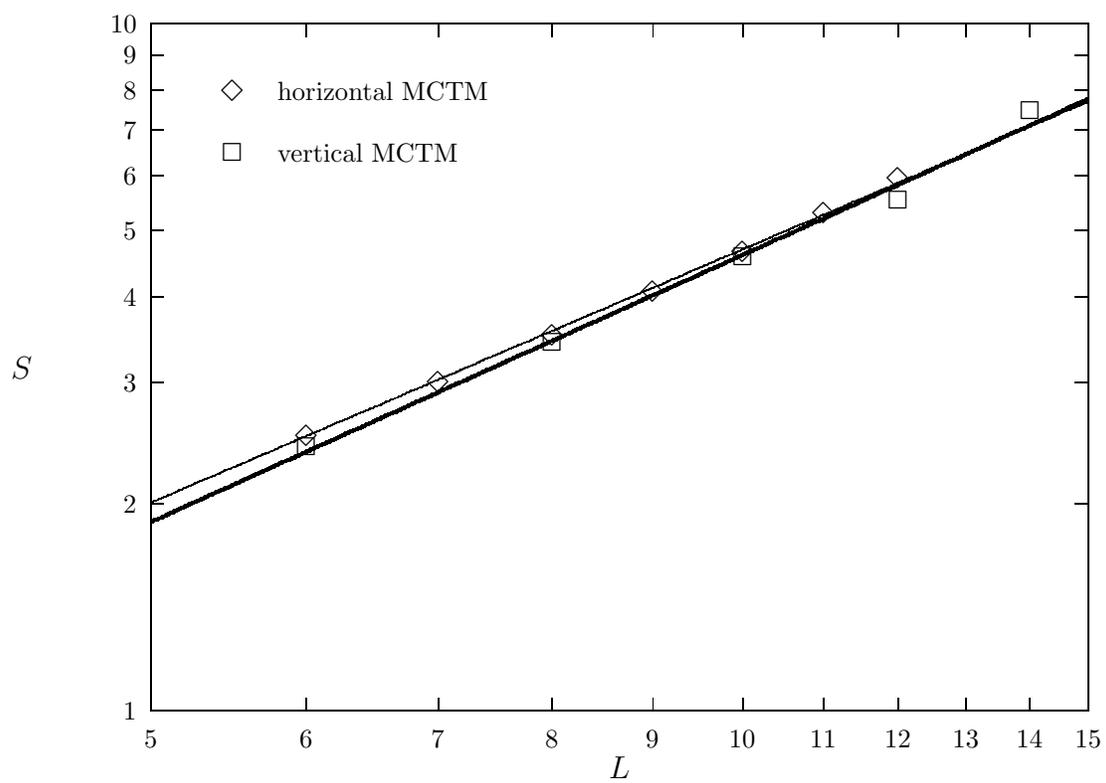
\begin{figure} \caption{$S=\partial \Delta F (K,L)/\partial K$ obtained
from  Monte Carlo transfer matrix data of
Figs.~\protect\ref{fig.DeltaFvsLh} and \protect\ref{fig.DeltaFvsLv}.
The exponent $1/\nu$ is obtained from the slopes of the linear fit of
horizontal (thin line) and vertical (thick line) Monte Carlo transfer
matrix calculations. }\label{fig.nu} \end{figure}

To determine the critical temperature and critical exponents, we make
use of the following finite-size scaling relation for the domain wall
free energy
\begin{equation}
 \Delta F(K,L) = A (L^{1/\nu} \Delta K),
\end{equation}
where $A$ is a scaling function and $\Delta K  =
K-K_{\rm c}$ is the deviation of the coupling constant from its
critical value $K_{\rm c}$.  For sufficiently small values of its
argument, the scaling function can be expanded as
\begin{equation}
\Delta F(K,L) = a_0 + a_1 L^{1/\nu} \Delta K + a_2 (L^{1/\nu} \Delta
K)^2 + \cdots,
\label{eq.scale}
\end{equation}
which shows that $\Delta
F(K,L)$ is constant as as function of $L$ for $K=K_{\rm c}$, but for
that value of the coupling constant only.  This behavior is apparent in
Figs.~\ref{fig.DeltaFvsLh} and \ref{fig.DeltaFvsLv} for at least $L \ge
8$ if we attribute the deviations for smaller system sizes to
corrections to scaling.  To the extent that the quadratic and higher
order terms in Eq. \ref{eq.scale} can be neglected, the critical
exponent $\nu$ can be obtained from the condition that $\Delta F(K,L)$
be linear in $L^{1/\nu}$ for fixed $\Delta K$.  In fact, in this
approximation \cite{Granato92}, $\nu$ can be obtained from the slope of
a log-log plot of $S= \partial \Delta F(K,L)/\partial K$ {\it vs} $L$,
which gives $1/\nu$ .  From the data of Figs.~\ref{fig.DeltaFvsLh} and
\ref{fig.DeltaFvsLv}, we have obtained $S$ as a function of $L$ for the
horizontal and vertical transfer matrix using this procedure. The
results are indicated in Fig.~\ref{fig.nu}.  The slopes of the straight
lines in the log-log plot, corresponding to the results for vertical
and horizontal transfer matrices, agree within the errors, as one would
expect, providing a check to the consistency of our data. From this
plot we can estimate $\nu = 0.80(4)$.  It is interesting to note that
within the linear approximation --as is the case for the free energy
barrier in the histogram of chirality in the MC simulations
\cite{LeeK}, where a similar finite size scaling is possible
\cite{LeeK90}  -- the estimate of the critical temperature $T_{\rm c} =
1/K_{\rm c}$ and the estimate of the thermal exponent $\nu$  are in
fact {\it independent} \cite{Granato92}, but this property is preserved only
approximately in our generalized finite-size analysis, where we fitted the
domain wall energies to the form of Eq.(\ref{eq.scale}) with as fitting
parameters $K_{\rm c},$ $\nu$, and $a_i$ with $i=0,1$, and $2$, as obtained by
truncation of the scaling function beyond second order. Note that the
correlation function exponent $\eta$ can be obtained from  the universal
amplitude $a_0$ in Eq.~\ref{eq.scale} using the results of conformal invariance
in two-dimensions \cite{Cardy}, $a_0 = \pi \eta$.

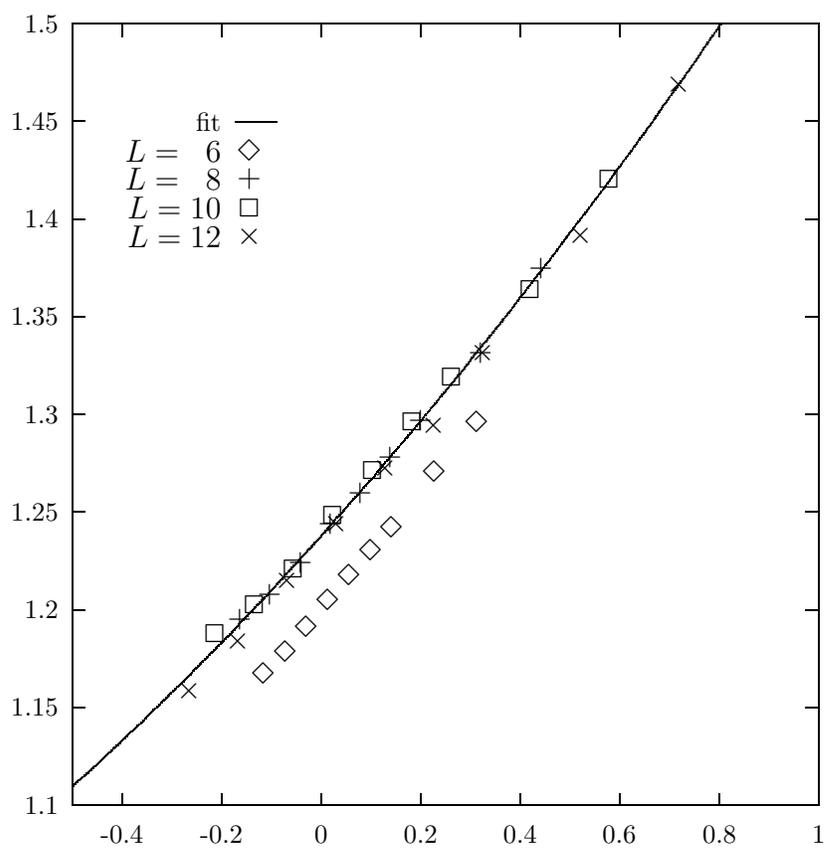
\begin{figure} \caption{Scaling plot of domain wall free energy data
obtained for the horizontal transfer matrix:  $\Delta F$ {\it vs}
$L^{1/\nu} \Delta K$.  The solid curve is the fitted scaling function
$A$, as given by the expansion in Eq.~\protect\ref{eq.scale}.  The data
for $L=6$ were not used for the fit.} \label{ScalingPlotHori}
\end{figure}

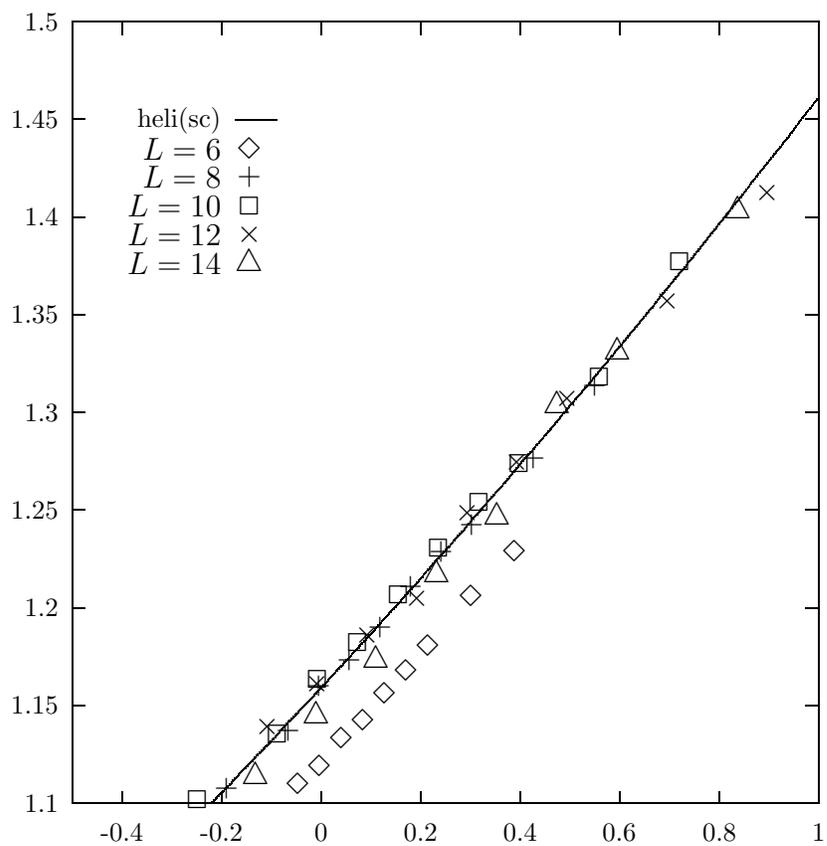
\begin{figure} \caption{Same as Fig.~\protect\ref{ScalingPlotHori} for
the vertical transfer matrix.} \label{ScalingPlotVert} \end{figure}

The results of the finite-size scaling based on Eq. \ref{eq.scale} fits
are summarized in Tab.~\ref{table.results} and the corresponding
scaling plots are shown in Figs.~\ref{ScalingPlotHori} and
\ref{ScalingPlotVert}.  The table contains estimates of the statistical
errors associated with the least-squares procedure.  In some cases the
$\chi^2$ were too large to be attributable to chance and as consequence
these statistical error estimates are to be treated with suspicion.
This is also evident from the discrepancies between the various
independent estimates.  Under the circumstances all we can do is to
take the mutual differences of various estimates as (admittedly
unsatisfactory) error estimates.  Thus we obtain our estimates: $T_{\rm
c}=0.454(3)$, $\nu=0.80(5)$, and $\eta=0.38(2)$. This result for $\nu$
is inconsistent with the pure Ising value of $\nu =1$, but it is in
agreement with the result $\nu = 0.85(3)$ from a similar finite size
scaling analysis of the free energy barrier in the histogram of
chirality from other Monte Carlo simulations \cite{LeeK}.  Our estimate
also agrees with a  recent estimate $\nu = 0.875(35)$ based on a
finite-size scaling analysis of correlation functions \cite{Jose}. We
therefore conclude that the estimate of $\nu = 1$ obtained by Thijssen
and Knops \cite{TK90} is likely to be an artifact of their fitting
procedure.

Recently, the phase diagram of the two dimensional XY-Ising model
defined by the Hamiltonian
\begin{equation}
 \beta H = - \sum_{<rr'>}[A(1+\sigma_r \sigma_{r'} )
 \cos (\theta_r - \theta_{r'}) + C \sigma_r
\sigma_{r'} ],
\label{eq.genXY}
\end{equation}
where $\sigma = \pm 1$,
has been studied in some detail \cite{Granato91,LeeG}.  This model is
expected to describe the critical behavior of a class of systems in
which  $U(1)$ and $Z_2$ symmetries play a simultaneous role, a class of
which the fully frustrated XY model is a special case. The critical
behavior of the square lattice fully frustrated XY model is represented
by the behavior of a particular (unknown) point in the parameter space
$(A,C)$ of this model. The phase diagram of model defined in Eq.
\ref{eq.genXY} consists of three branches, in the ferromagnetic region,
joining at $C \approx 0$. Along one of these branches ($C < 0$) the
transition takes place at a critical point with simultaneous
criticality of $U(1)$ and $Z_2$ order parameters.  Along this line, the
critical behavior appears to be non-universal with $\nu$ varying from
0.76  to 0.84 and $\eta$ varying from  0.25 to 0.5 along a segment of
the line.  Further away from the branch point, this line of single
transition appears to become first order.  Our results for the chiral
critical exponents  of the square lattice fully frustrated XY obtained
by the Monte Carlo transfer matrix method do in fact appear to be
consistent with the critical exponents of a particular point along the
line of continuous, single transitions of the XY-Ising model. Once
this  notion is accepted, it implies that the fully frustrated XY has a
single nonuniversal transition. The non-universality of this transition
in turn suggests that the critical exponents of the square and
triangular lattice fully frustrated XY models may differ from each
other, although they may be quite close \cite {LeeK}.

Besides critical exponents, another important quantity that can be
inferred from Monte Carlo transfer matrix calculations, is the central
charge $c$, which classifies the possible conformally invariant
critical theories \cite{Blote}. The central charge is related to the
amplitude of the singular part of the free energy per site, at
criticality, in the infinite strip by
\begin{equation}
 f(K_{\rm c},L) = f_\infty + {{\pi c}\over {6L^2}}
\end{equation}
which
is valid asymptotically for large $L$. Fitting  the data for $f(K,L)$
closest to the estimated critical temperature $T_{\rm c}$, we obtain
$c=1.61(3)$ from the strips of $ 8 \le L \le 14$ using both horizontal
and vertical Monte Carlo data. This result agrees with the estimate of
the central charge first obtained by Thijssen and Knops \cite{TK90},
also using Monte Carlo transfer matrix calculations, and appears to be
significantly larger than $c=3/2$, which would be expected if the
transition was single, but decoupled \cite{Foda}.  However, one cannot
be certain of this value unless calculations are done at sufficiently
large $L$ such that small-$L$ corrections to the above asymptotic
expression are negligible. Using only the Monte Carlo transfer matrix
data  for $ 6 \le L \le 14$, may not allow us to extrapolate to the $L$
large limit and it is quite likely that this estimate of $c$ is subject
to systematic errors. In fact, recent large $L$ calculations for the
related XY-Ising model \cite{NightingaleG} show a significant decrease
of $c$ with increasing $L$. Unfortunately, our data for fully
frustrated XY model for $L > 14$ turn out to be rather noisy. This
problem can be remedied in principle by using variance reductions
techniques
\cite{NightingaleCaflisch88,NightingaleSuper88,NightingaleRev90}, but
as discussed this is complicated in the case of the square lattice
fully frustrated XY model because of the presence of both ferromagnetic
and antiferromagnetic bonds. The case of the triangular lattice is more
straightforward in this respect.

\section{Conclusion}

We have studied the finite-size behavior of the chiral domain wall free
energy of the square lattice fully frustrated XY model on an infinite
strip, using the Monte Carlo transfer matrix. From a finite-size
scaling analysis of data for strip widths up to $14$ lattice spacing,
we have estimated the  critical temperature and chiral critical
exponents.  The latter appear to be significantly different from the
pure Ising values and in agreement with other recent Monte Carlo
simulations. The results are also consistent with the identification of
the critical behavior of the fully frustrated XY model with that of a
point of the line of single transitions in XY-Ising model studied
recently \cite{Gfootnote}.  The value of the central charge $c=1.61(3)$
is found to be
consistent with the estimate first obtained by Thijssen and Knops, but
we cannot rule out the possibility that all results obtained for this
and similar models are skewed by large, slowly decaying
corrections-to-scaling.

\section{Acknowledgments}
 We thank J.M. Kosterlitz, T. Ala-Nissila and K. Kankaala for many
helpful discussions. Monte Carlo sampling of the XY degrees of freedom
was performed with a routine developed in a previous collaboration with
Dr. H.W.J.  Bl\"ote.  Part of the computations were carried out in a
Cray XMP with computer time provided by the Scientific  Computational
Center of Finland and their generous support is gratefully
acknowledged. This work was supported by Funda\c c\~ao de Amparo \`a
Pesquisa do Estado de S\~ao Paulo (FAPESP, proc. no. 92/0963-5)
and Conselho Nacional de Desenvolvimento Cient\'ifico e Tecnol\'ogico
(CNPq) (E.G.)
and by the National Science Foundation under grant numbers DMR-9214669
and CHE-9203498(M.P.N.).

\newpage \begin{table} \caption{ Results for critical temperature and
critical exponents, obtained from finite-size scaling analysis of
domain wall energies.  Standard errors are indicated parenthetically.
As explained in the text, these standard error reflect only statistical
uncertainties, which are presumably considerably smaller than the
errors due to corrections to scaling.  The column labeled $L$ indicates
which system sizes were used in the fits: $L_1,L_2\ (\Delta L )$ stands
for sizes from $L_1$ to $L_2$ in steps of $\Delta L$. Horizontal and
vertical transfer matrix are indicated by v and h under the heading
transfer. The last column is the $\chi^2$ per degree of freedom.}

\vskip 1cm


\end{center}
\end{figure}

\centerline{} \vfill \centerline{Fig.~\ref{ScalingPlotVert} }


\begin{references}
\bibitem{TKfootnote} The analysis used by Thijssen and Knops
\cite{TK90} consists in extrapolating the finite-size data to the
thermodynamic limit and subsequently extracting $\nu$ from the
singularity of the extrapolated function.  The problem of this approach
is that there is no reliable procedure to obtain a function which is
singular in the thermodynamic limit as the extrapolation of regular
functions computed for finite systems.  Procedures using finite-size
scaling relations such as the ones we employ avoid this problem.
\bibitem{Gfootnote} Essentially the same results have also been obtained
for a one-dimensional quantum version of the fully frustrated
XY model which can describe  the $T = 0$ superconductor-insulator transition
due to charging effects in a ladder of Josephson junctions arrays at
a half a flux quantum per plaquette \cite{Granato92}.
\end{references}
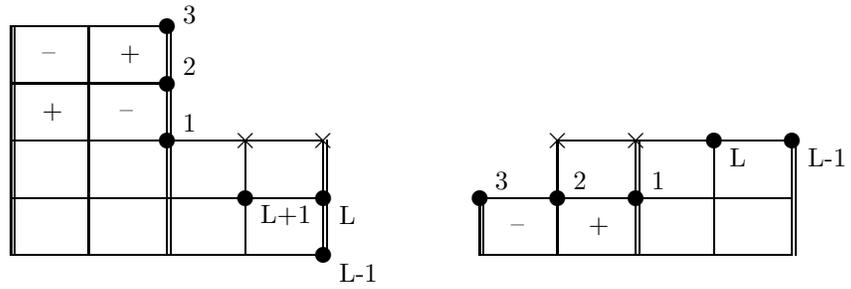
\end{document}